# Coded Path Protection:
# Efficient Conversion of Sharing to Coding

Serhat Nazim Avci and Ender Ayanoglu
Center for Pervasive Communications and Computing
Department of Electrical Engineering and Computer Science
University of California, Irvine

*Abstract*—Link failures in wide area networks are common and cause significant data losses. Mesh-based protection schemes offer high capacity efficiency but they are slow and require complex signaling. Additionally, real-time reconfiguration of a cross-connect threatens their transmission integrity. On the other hand, coding-based protection schemes are proactive. Therefore, they have higher restoration speed, lower signaling complexity, and higher transmission integrity. This paper introduces a coding-based protection scheme, named Coded Path Protection (CPP). In CPP, a backup copy of the primary data is encoded with other data streams, resulting in capacity savings. This paper presents an optimal and simple capacity placement and coding group formation algorithm. The algorithm converts the sharing structure of any solution of a Shared Path Protection (SPP) technique into a coding structure with minimum extra capacity. We conducted quantitative and qualitative comparisons of our technique with the SPP and, another technique, known as *p*-cycle protection. Simulation results confirm that the CPP is significantly faster than the SPP and *p*-cycle techniques. CPP incurs marginal extra capacity on top of SPP. Its capacity efficiency is lower than the *p*-cycle technique for dense networks but can be higher for sparse networks. In addition, unlike *p*-cycle protection, CPP is inherently suitable for the wavelength continuity constraint in optical networks.

## I. INTRODUCTION

Studies show that reasons of failure in networks can be widespread. According to [1], cable cut rate per 1000 sheath miles per year is 4.39. That means on average a cable-cut occurs every three days per 30,000 fiber miles. These numbers are consistent with the FCC data, summarized as 13 cuts per year for every 1,000 miles of fiber and 3 cuts per year for every 1,000 miles of fiber for metro and long haul networks respectively [2]. As stated in [3], 70% percent of the unplanned network failures affect only single links. For this reason, in this paper, we focused on single link failure recovery.

1+1 and 1:1 automatic protection switching were early attempts of path-based protection mechanisms but were dropped due to low capacity efficiency. Mesh-based protection schemes attracted attention due to their high capacity efficiency but suffered from low speed. SPP [4] is a widely recognized mesh-based path protection technique. It specifies two link-disjoint paths for each connection and reroutes the traffic over the protection path if the primary path fails.

Reference [5] introduced the concept of a *p*-cycle in order to achieve both fast restoration and low spare capacity percentage. Fundamentally, a *p*-cycle is a mixture of ring-type protection and link-based protection. Its performance is similar to SPP in terms of resource utilization and similar to link-based protection in terms of restoration time. In the case of a failure in a link protected by the cycle, the affected traffic is rerouted over the spare capacity in the healthy parts of the *p*-cycle.

The *p*-cycle approach achieves higher restoration speed by simply minimizing the number of optical cross-connect (OXC) configurations after failure. "Hot-standby" [6] and "pre-cross-connected trials" (PXT) [7], which are extensions of SPP, are developed based on the same idea. We offer a novel proactive protection scheme called Coded Path Protection (CPP). It is faster and more stable than rerouting based schemes because it eliminates the real-time OXC configurations after failure. The capacity placement algorithm of CPP is based on converting the sharing operation of SPP into coding and decoding operations with a slight extra cost. Integer linear programming (ILP) is incorporated to carry out the optimal conversion with minimum total capacity. In the next sections, comparisons between our schemes and aforementioned conventional techniques are performed, Simulations over realistic network scenarios using ILP formulations are carried out.

## II. RELATED WORK

The idea of incorporating network coding into link failure protection dates back to 1990 [8] and 1993 [9], prior to the first papers on network coding [10]. The technique is called *diversity coding,* and in it, $N$ primary links are protected using a separate $N+1^{st}$ protection link which carries the modulo-2 sum of the data signals in each of the primary links. If all of the $N+1$ links were disjoint or physically diverse, then any single link failure could be recovered from by applying the modulo-2 sum over the received links. Assume that data on the primary links are $b_1, b_2, b_3, \ldots, b_N$ and the checksum of the primary data is

$$c_1 = b_1 \oplus b_2 \oplus \cdots \oplus b_N = \bigoplus_{j=1}^{N} b_j.$$

In the receiver side of the operation, if a failure is detected, the decoder applies modulo-2 sum to the rest of the $N$ links and extract the failed data as

$$c_1 \oplus \bigoplus_{\substack{j=1 \\ j \neq i}}^{N} b_j = b_i \oplus \bigoplus_{\substack{j=1 \\ j \neq i}}^{N} (b_j \oplus b_j) = b_i$$

where we assumed $b_i$ is the failed link. This operation is fundamentally different than rerouting-based protection schemes since it does not need any feedback signaling.

This idea was revisited by the authors of this paper in [11] and a coding structure for an arbitrary network topology was developed. This scheme may require extra links from the destination nodes to decoding nodes to be able to decode signals. It has been shown in [11] that diversity coding can achieve higher capacity efficiency than the SPP and the *p*-cycle techniques.

In [12], a bidirectional protection scheme that uses network coding over *p*-cycle topologies on mesh networks was introduced and called as 1+N protection. The idea presented in [12] is to form circular protection paths in both directions that traverse over the source and destination nodes of the group of flows that are to be protected. In [13], a new tree-based protection scheme was introduced instead of a *p*-cycle based scheme and called Generalized 1+N protection (G1+N). In [13], same data from both end-nodes are sent on a parity link. Symmetric transmission is broken only for the connection affected from the failure. The capacity efficiency of G1+N is basically unknown. However, it clearly lacks the speed of diversity coding since the distance between the destination nodes and the decoding node is much longer than for diversity coding. In [14], a new trail-based protection scheme is proposed. Failed data is recovered via a linear coded protection circuit. This structure is a modified version of the scheme in [12] resulting in higher capacity efficiency by moving from cyclic to linear protection topology.

## III. CODED PATH PROTECTION

In this paper, we propose a novel coding technique, which we call Coded Path Protection (CPP). We present a simple strategy to find the optimal coding structure without much complexity. Coded path protection is faster, has less signaling complexity, and has higher transmission integrity than any of the rerouting-based protection techniques. Spare capacity percentage (SCP) of coded path protection is slightly larger than the SCP of shared path protection. Our contribution in this paper consists of two parts, namely a novel coding structure and a simple but optimal coding group formation algorithm.

The set of operations starting with the input of network and traffic data until reaching a CPP solution with a valid coding structure are shown in Fig. 1 in a sequential order. On the contrary, in this section, we start with explaining our methodology in converting a typical solution of Shared Path Protection into a one with sharing replaced by coding, which we call Coded Path Protection. Then, we show how to establish a valid coding structure and show that the encoding and decoding inside the network can be carried out within this coding structure. In the next section, first, we explain the design algorithm that finds a Shared Path Protection (SPP) solution to work on. Second, we present a design algorithm, which optimally converts the sharing structure into the coding structure given the solution of SPP.

We benefit from the basic coding structure of [14] while building a valid coding structure in CPP. When the traffic is bidirectional, the end-nodes of a connection generate the same set of protection signals and transmit them over the protection path to the other end-node of the connection so that the failed data can be recovered from the protection path shortly after the failure. This proactive protection mechanism makes 1+N coding [14] faster than the sharing based protection schemes. This structure creates a symmetricity over the protection path of each bidirectional connection. The parity data is formed by applying XOR operation to the data received from and the data transmitted over the primary path. Encoding and decoding of different parity data inside the network can be done by utilizing the symmetricity over the protection paths of the connections. Despite its restoration time advantage, 1+N coding works on a specific limited linear topology, which is why it falls short in exploring the full connectivity inside the network. On the other hand, CPP mostly preserves the topology of an SPP solution with a small compromise on connectivity without losing the speed advantage that comes inherent from the coding structure.

Symmetric transmission is key in coding and decoding operations. This is illustrated in Fig. 2(a) in an example with two connections. Thick straight lines are primary paths and dotted lines are protection paths. For the time being, synchronization and timing are not considered. Assume that $S1$ transmits $s_1$ to $D1$ and $D1$ transmits $d_1$ to $S1$ using the primary path at time $t_0$. After a delay of $\tau$, these signals are received by the reciprocal nodes at the same time and both end-nodes form the summation of these two signals, mathematically $c_1 = s_1 \oplus d_1$. At time $t_0 + \tau$, the same $c_1$ symbols are sent from the corresponding end-nodes of the protection path of $S1 - D1$. It is similar for $S2 - D2$. As it is seen at Fig. 2(a), $c_1$ and $c_2$ are coded over the link $A - B$. $A$ is the node where $c_1$ and $c_2$ are coded and node $B$ is responsible for decoding. Node $B$ extracts $c_1$ using $c_1 \oplus c_2$ and $c_2$, and extracts $c_2$ using $c_1 \oplus c_2$ and $c_1$. Therefore coding and decoding of signals $c_1$ and $c_2$ are successfully completed and node $B$ transmits them to $D1$ and $D2$ over links $B - D1$ and $B - D2$ respectively. Fig. 2(b) gives an example of single link failure on the primary path of $S1 - D1$. Due to failure, $S1$ receives 0 instead of $d_1$ and transmits $s_1 \oplus 0 = s_1$ at time $t_0 + \tau$ over the protection path. Similarly, $D1$ receives 0 instead of $s_1$, so it transmits $0 \oplus d_1 = d_1$. These signals are coded with $c_2$ over link $A - B$ and they are decoded at nodes $B$ and $A$ respectively. At node $B$, $s_1$ is extracted by summing $s_1 \oplus c_2 \oplus c2 = s_1$ and forwarded over the link $B - D1$. At node $A$, $d_1$ is acquired by summing $d_1 \oplus c_2 \oplus c_2 = d_1$ and forwarded over the link $A - S1$. Reciprocity over the protection path of $S2 - D2$ enables perfect coding and decoding of the data signals of the failed connection.

"Poison-antidote" analogy [15] is useful in understanding the general coding structure. When two signals are coded together they "poison" each other. At the decoding node, "antidote" data are needed to extract the signals from each other. For the general two connection case in Fig. 3, same signals traverse the reverse directions over the protection path. Straight lines are protection paths of $a$ and $b$ in one direction. Dotted lines are protection paths of same $a$ and $b$ in the reverse direction. At node $A$, straight paths are antidotes of dotted paths. At node $B$, dotted paths are antidotes of straight paths. In the single link failure case, if connections have link-disjoint primary paths, at most one of them is affected from the failure. The other connection can preserve reciprocity and the poison-antidote structure to help recover the affected connection.

We can generalize this coding structure to an arbitrary number of connections, arbitrary number of links, and to an arbitrary topology by the use of reciprocity. However, first

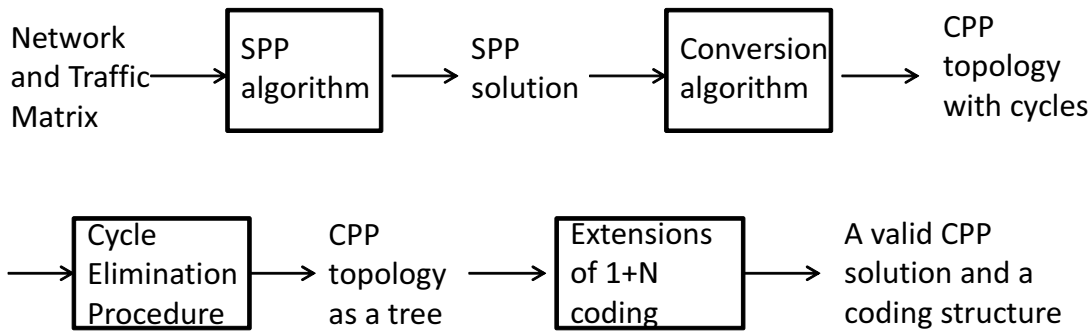

Fig. 1. Steps in creating a CPP solution.

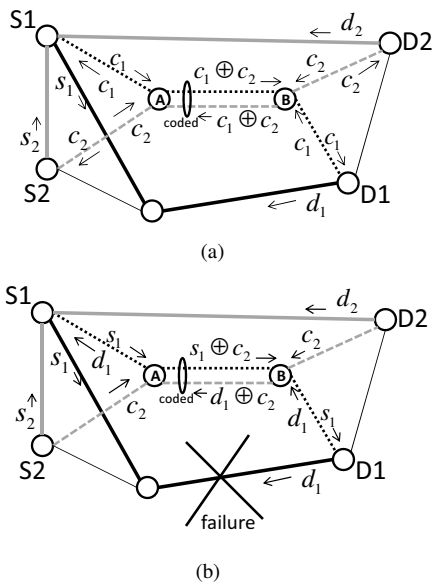

(a)

(b)

Fig. 2. Coding and decoding operations for coded path protection, (a) In normal state, (b) Reaction to single link failure

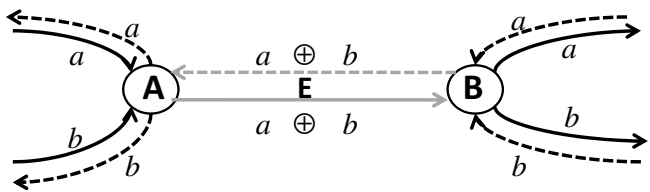

Fig. 3. Coding at an arbitrary link and decoding at an arbitrary node.

we should define the concept of a coding group. Assume that $S1-D1$ and $S2-D2$ are coded over some link $E$. Then, they are considered to be in the same coding group. This group gets bigger if either $S1-D1$ or $S2-D2$ is coded with some other connection. For example, if $S3-D3$ is coded with $S2-D2$, it is also considered in the same coding group with both $S2-D2$ and $S1-D1$. In a coding group, coding structure can recover from a single link failure on one of the primary paths if the reciprocity property is preserved for the other connections. To guarantee this property, two protection paths can be coded together as long as

1) Their primary paths are link-disjoint,
2) Their primary paths are also link-disjoint with the primary and protection paths of the connections in the same coding group.

These are sufficient link-disjointness rules to satisfy the decodability on arbitrary CPP topologies. In III-D, these rules can be relaxed to some extent to utilize network connectivity more. These rules can also be interpreted as the criteria of two connections to be in the same coding group. In addition to these rules, the primary and protection paths of the same connections are inherently link-disjoint as a cardinal rule in path-based link failure recovery.

The scheme in [14] is similar to CPP with two differences. First, its capacity efficiency is not known. Second, there does not exist a simple and optimal algorithm to implement it for an arbitrary topology. However, CPP is suitable to convert a typical solution of SPP with low complexity because a typical solution of SPP must obey the first rule above. The rest of the work to convert an SPP solution to CPP is to form coding groups that satisfy the second rule.

We assume that for a given topology and a given set of connections, there is a pre-calculated solution of SPP. Given the solution, primary and protection paths of the connections, wavelength assignments, and maximum required spare capacity on each link will be known. Referring to Fig. 4(a), thick straight lines represent the primary paths of end-to end connections, whereas protection paths are stated by dotted lines. In Fig. 4(a), numbers associated with edges are index values of edges. Some of the protection capacity is shared by multiple protection paths. There is a limited freedom in terms of choosing the group of connections which will share the same capacity over the same link. For example, $S3-D3$ can share the one unit spare capacity at link 5 either with connection $S1-D1$ or with connection $S4-D4$. However, $S1-D1$ and $S4-D4$ cannot share that capacity since their primary paths are not link-disjoint. This freedom can be utilized in converting sharing groups to valid coding groups with zero or unappreciable additional capacity.

In the given solution of SPP, protection paths are coupled under the provision of the first rule. However, while building the CPP solution, protection paths are coupled and coding groups are formed in a way such that both rules are satisfied. The sharing structure in Fig. 4(a) is converted to the coding structure in Fig. 4(b) in this manner. It should be noted that at link 5 $S1-D1$ and $S3-D3$ are coupled to share the one unit capacity in the SPP solution. However, in CPP solution $S3-D3$ is coded with $S4-D4$, not with $S1-D1$. If that is not done, then $S1-D1$, $S2-D2$, $S3-D3$ and $S4-D4$ would be in the same coding group because they would be indirectly related. Then the second rule about link-disjointness would not be satisfied. After this modification, we can divide this coding group into two, one group consists of $S1-D1$ and $S2-D2$ and the other consists of $S3-D3$ and $S4-D4$. Then both of the rules are satisfied. In this example, no extra capacity is required to convert a SPP solution to a CPP solution with the aid of limited freedom in the SPP solution. However, that is not the case in general. Therefore, we developed an ILP formulation to conduct the conversion with minimum extra capacity.

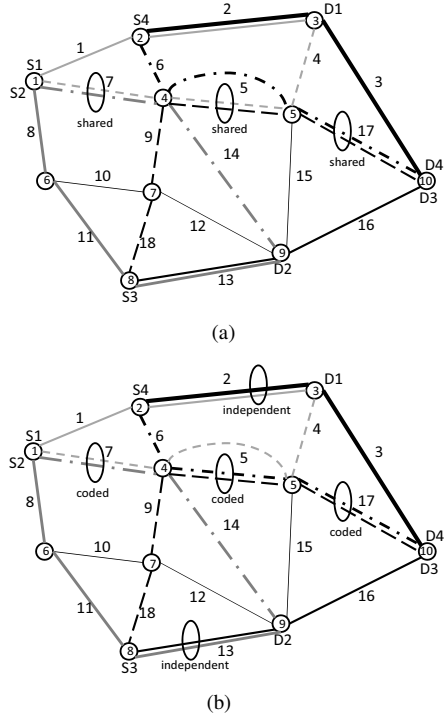

(a)

(b)

Fig. 4. Possible coding and sharing scenarios over a network, (a) Sharing of protection capacities, (b) Coding of protection paths

### A. Cycle Elimination

The outputs of the conversion algorithm are the coding group combinations and the protection topologies of each coding group. Even the conversion algorithm is optimal, there can occur some cyclic structures inside some of these topologies, which cause suboptimality. In [13], by Proposition 1, it is shown that *"under the assumption of undirected edges in the network graph G, the minimal cost protection circuit, $P_i$, where the cost is in terms of the number of network edges, is a tree."* The suboptimality inside the CPP topologies occurs due to two reasons. First, the design algorithm of SPP may not be optimal due to high complexity. Second, the groupings under the sharing configuration is rearranged by the conversion algorithm, which may lead to some cyclic structures in some of the coding group topologies. We eliminate these cyclic structures to transform the coding group topologies into tree structures, which is critical in building the coding structure of CPP. The elimination of the cyclic structures is shown in Fig. 5 and they lead to further capacity savings. The signals in the longest link of the cyclic structure is coded with the signals on the rest of the cycle, shown with dashed links. Therefore, the capacity of the longest link is saved. The link-disjointness rules ensure that the rest of the cycle does not share any link with the primary paths of the connections on interest.

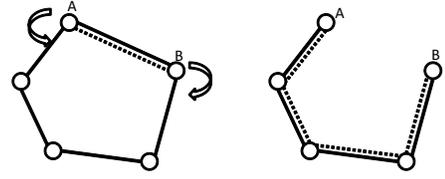

Fig. 5. Removal of a cyclic structure

### B. CPP Coding Structure

We need to prove that simple linear coding structure of 1+N coding can be extended to any arbitrary tree structure in order to implement this idea over arbitrary CPP protection topologies, which are tree-like. Before demonstrating how to build general coding structure of CPP, we need to show the extensions that can be done over the simple linear coding structure of 1+N coding. The basic structure of 1+N coding protection circuit (trail) for 3 connections is shown in Fig. 6. The link-disjoint primary paths between end-nodes are not shown for clarity.

*1) Lemma 1:* In the linear topology of 1+N coding, a node can serve as the end-nodes of multiple connections. In this case, these end-node can be represented by separate hypothetical adjacent end-nodes on the linear coding graph (trail). The links between these end-nodes are assumed to have zero length. In other words, multiple end-nodes over the linear coding structure may refer to same physical node if the links between them has zero length. Each end-node can be separated from each other since they are connected to the physical node via independent ports as shown in Fig. 7(a). The parallel horizontal links represent the coding trail passing through the nodes of interest. Let $E$ be the set of end-nodes which share the same physical node $F$. Since the information regarding each end-node is independent from each other, they can be separately depicted with the hypothetical end-nodes in Fig. 7(b).

*2) Lemma 2:* The classical 1+N coding requires each end-node to be traversed by the common protection path. However, the same coding structure can be applied even if an end-node is connected to the linear topology through a direct path which deviates from the common trail. In Fig. 8(a), the end-node $T_i$ is connected to the linear coding topology via an arbitrary on-trail node $D$ through a bidirectional link. In terms of coding operation at node $D$, there is no change if the node $D$ and node $T_i$ are assumed to be the same node. This transformation is depicted in Fig. 8(b), where the dashed box combines these two nodes into a single one in terms of coding operations over the linear topology. In Fig. 8(c), Fig. 8(b) can simplified and node $T_i$ can represented on the linear coding trail via a hypothetical node $T'_i$. We can generalize this operation to any arbitrary number of end-nodes as long as they are connected to common trail via link-disjoint paths.

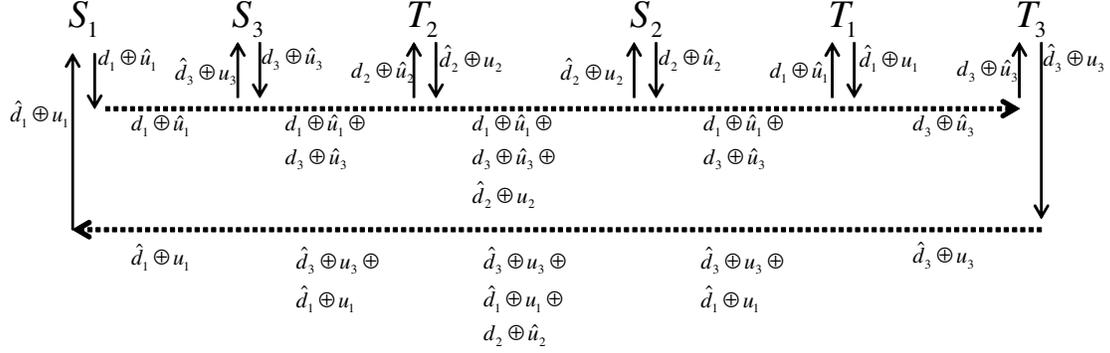

Fig. 6. Coding at 1+N protection [14] circuit for 3 connections.

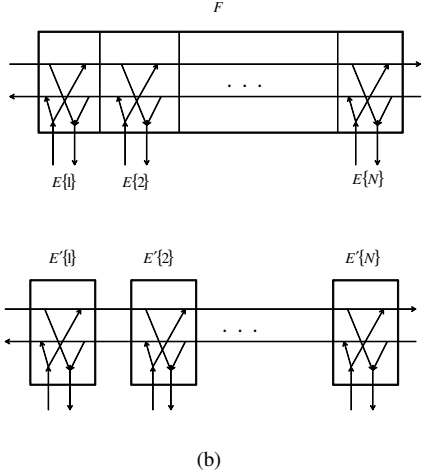

Fig. 7. Proof of Lemma 1 (a) Multiple end-nodes share the same node, (b) Each end-node can be shown as a separate entity over the protection trail

*3) Lemma 3:* As an extension to the second *Lemma*, if $N$ number of end-nodes are connected to a node on the trail via a common link, these end-nodes can be still represented over the trail by a different notation. This is useful if the separate signals of these end-nodes are not the interest. In Fig. 9(a), end-nodes $S_i$ and $T_j$ are combined at an arbitrary node $C$ and $C$ is connected to an arbitrary node $D$ over the linear coding trail. In this case, the the end-nodes cannot be represented independently because there is no mechanism to decode the signals in node $C$. However, from the network point of view, these two end-nodes can be merged into a single end-node as $S_i \oplus T_j$ since the coding operations in the rest of the network is not affected. Note that $S_i \oplus T_j$ is only a notation because the end-nodes cannot summed but their parity signals are summed. The new hypothetical node is depicted in Fig. 9(b). In Fig. 9(c), the node $S'_i \oplus T'_j$ is hypothetically placed over the trail using second *Lemma*. The number of combined end-nodes can be set an arbitrary number $N$ and the hypothetical end-node will be represented as the summation of all the combined end-nodes.

*4) Lemma 4:* If we merge any arbitrary number of adjacent end-nodes over the linear coding trail, the coding operations in the rest of the trail is unaffected. Let $P$ be the set of adjacent end-nodes which are supposed to be merged into a single end-node over the trail. In Fig. 10(a), the coding and decoding operations inside these end-nodes are shown. This structure can be converted to the structure in Fig. 10(b), in case the individual signals of the end-nodes in $P$ are not necessarily extractable. Then, the combination of these end-nodes is represented with a single hypothetical end-node as shown in Fig. 10(c).

*5) Lemma 5:* If the extensions to the linear coding trail does not create a cyclic structure inside the topology, it is possible to separate the topology into systems. In Fig. 11(a), the coding topology can be divided into 2 subsystems with the dashed link. These subsystems are highlighted in Fig. 11(b). One of them is the hypothetical end-node, defined by *Lemma 4*, represented by the summations of multiple end-nodes, which are spanned by a common link from the linear coding trail. The other subsystem is the rest of the coding topology, which is the rest of the tree. Regarding the input and output relationship between these two subsystems at that specific branch point $D$, it is seen that these subsystems are the complementary of each other. The complementary hypothetical end-node is formulated in Fig. 11(b).

*6) Example 1:* In order to visualize how these lemmas are useful in transforming a tree topology into a linear trail topology, an example is provided below. Assume that, there are 6 bidirectional connections such that $S_i$ is communicating with $T_i$ for $i = 1, 2, 3, 4, 5, 6$. There exist a bidirectional primary path between each end-node pairs which is link-disjoint to the other primary paths and to the common protection trail. In Fig. 12(a), the end-nodes of the connections are shown on the network. For clarity, link-disjoint primary paths are not depicted. The dashed link is the linear coding trail which has the coding structure of 1+N coding. The protection paths has a tree-like topology. Using the *Lemmas 1 to 3* defined previously, we can convert this tree-like topology into a trail coding topology. The hypothetical nodes are highlighted with a prime sign. They are no different than regular end-nodes in terms of coding operations over the trail. The converted structure is given in Fig. 12(b). The end-nodes, which are shown as single entities over the trail, can successfully extract their parity data from the trail. In case of a failure in their primary paths, they can recover the failed data from the trail as shown in [12].

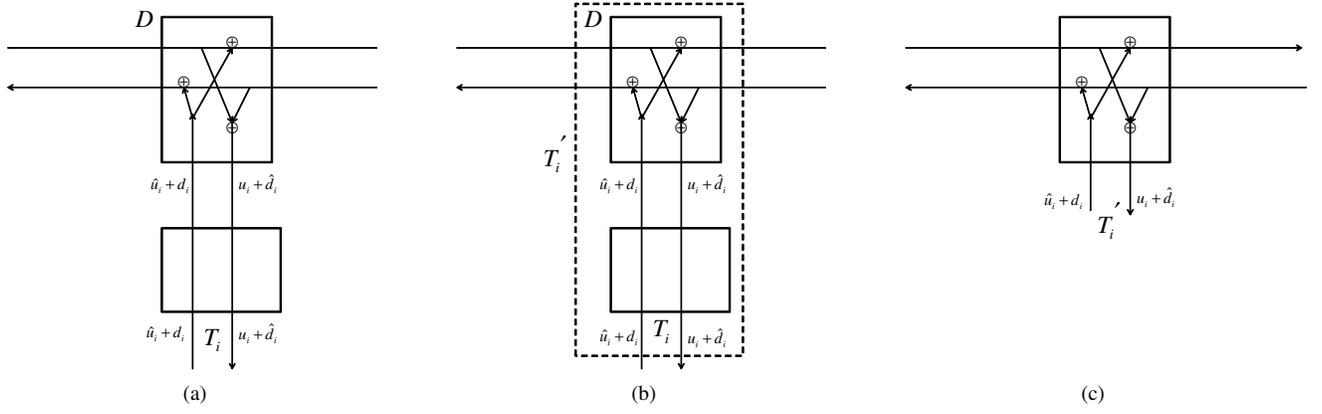

Fig. 8. Proof of Lemma 2 (a) An end-node is connected to the coding trail via a direct path, (b) From trail point of view, they are seen as a single node, (c) The end-node hypothetically is over the trail

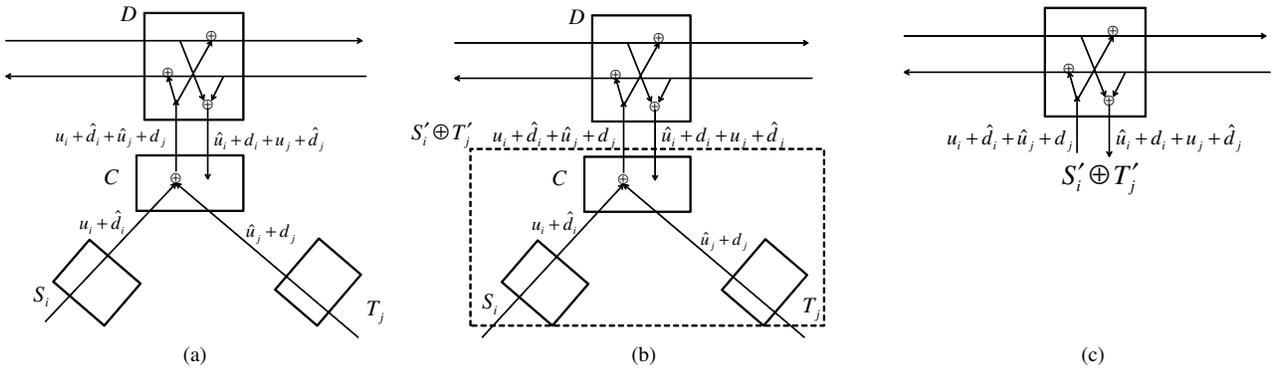

Fig. 9. Proof of Lemma 3 (a) Two different end-nodes are connected to the trail via the same link, (b) They can be merged into a single node, (c) How they are seen from the rest of the trail

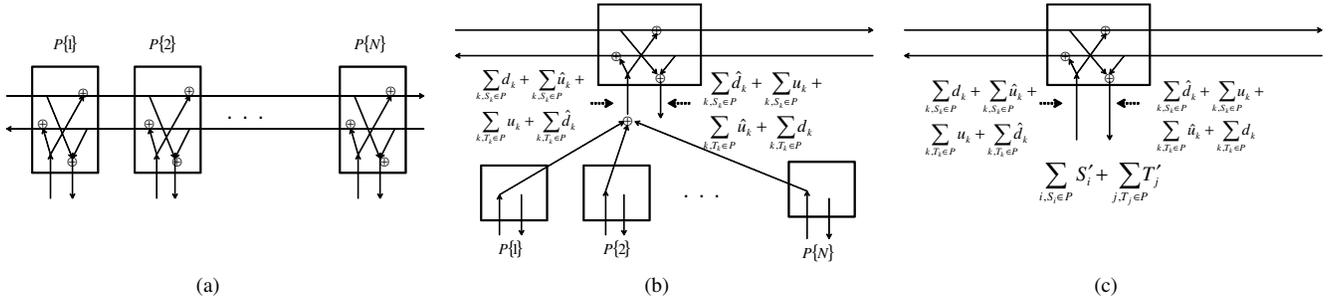

Fig. 10. Proof of Lemma 4 (a) Multiple adjacent end-nodes over the trail are depicted, (b) These end-nodes can be merged into a single one, (c) How they are seen from the rest of the trail.

In the next step, some of the adjacent end-nodes are merged. In Fig. 12(c), the end-node pairs $S_2 - T_4$ and $T_3 - S_5$ are merged into single hypothetical nodes using *Lemma 4*. In Fig. 12(d), at the specific branch point $D$ over the trail, the coding topology is divided into two subsystems. The underlying topologies are shown inside the boxes. The notations outside the boxes are the images of the subsystems as they are seen from the opposite subsystem.

### C. CPP Coding Structure

Assume that one of the coding groups in CPP solution has $N$ connections and the end-nodes are given in the set $P = \{S_i, T_i : 1 \leq i \leq N\}$ meaning that each connection consists of the end-nodes with the same indices. As stated before, the protection topology of this coding group is link-disjoint to the primary paths in the same coding group and it is a tree. The end-nodes of the connections are scattered over this tree. A valid encoding and decoding structure is established using the

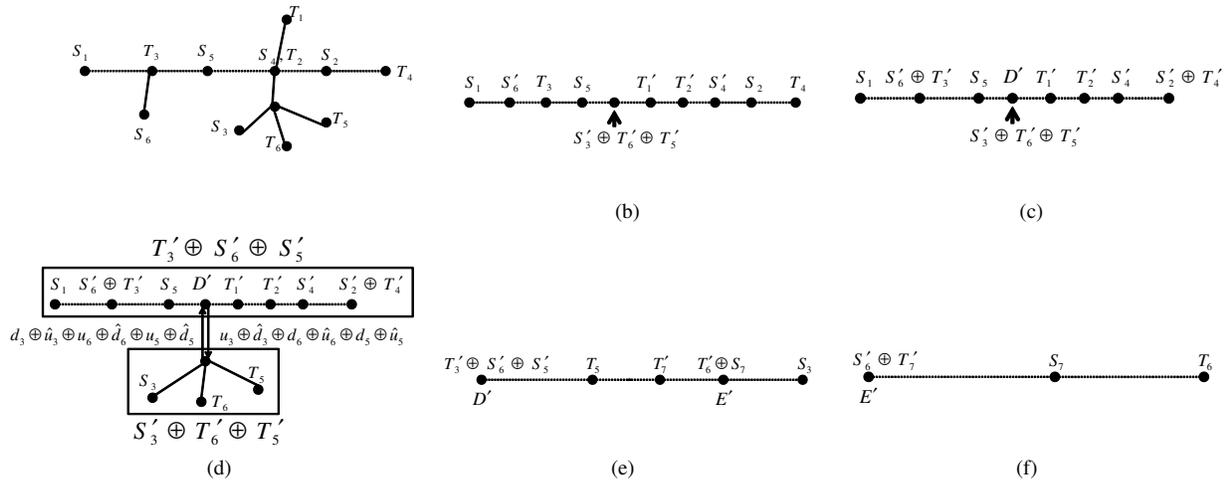

Fig. 12. Conversion from a tree to a linear coding trail (a) The tree-like topology of the protection paths, (b) The end-nodes are represented over the trail, (c) $D$ is the branch trail and some end-nodes are merged (d) The tree can be divided into two separate subsystems, (e) the branch trail originating from node $D$ and $S7 - D7$ is introduced, (f) The sub-branch trail originating from node $E$ of the branch trail

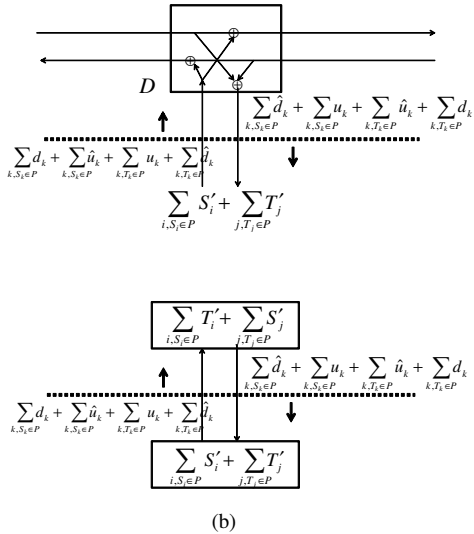

Fig. 11. Proof of Lemma 5 (a) Input-output relationships between the end-node and the rest of the coding trail, (b) The rest of the trail is treated as the combination of some of the end-nodes

following steps.

1) Select one of the links inside the tree and call it the truck trail.
2) Extend this truck trail from both ends as long as the extended links reach to the edges of the tree. When there are multiple links to extend, one of is randomly selected.
3) When the truck trail reaches its limits, using *Lemmas 1 to 3*, place the end-nodes over the trail. There are three types of end-nodes. The end-nodes which are physically over the trail are shown as separate entities over the trail with the help of *Lemma 1*. The second type of end-nodes are not physically over the trail but directly connected to the trail via a dedicated path. They are depicted over the trail with the help of *Lemma 2*. The third type of end-nodes are connected to the trail via a common link or common links. These end-nodes are placed over the trail as a combination of multiple end-nodes with the help of *Lemma 3*. We call the hypothetical nodes which represent the combination of multiple end-nodes as the branch points on the trail. There can be multiple branch points over a single trail.
4) Assume $R$ is the set of combination of multiple end-nodes as $R = \{R_1, R_2, ..., R_k\}$, where $k$ is the total number of branch points over the truck trail. Each $R_i$ keeps the end-nodes that are spanned by the branch point $i$. If the same pair of end-nodes belong to the same set $R_i$, $1 \leq i \leq k$, omit them from the truck trail. They will be taken into account later.
5) Then, code the signals whose end-nodes are over this truck trail as it is explained in [12] under 1+N protection coding operations. The truck trail is the protection circuit of 1+N coding. The end-nodes which are shown as single entities will be able to receive their parity data from the trail. In case of failure, these nodes are able to extract the failed data from the linear 1+N coding trail.
6) The remaining end-nodes are the ones who are depicted as the combination of multiple end-nodes. There are $k$ combinations and each combination has a branch point. Originating from these branch points, new branch trails will be initiated using the links that spans the end-nodes in sets $R_i$, $1 \leq i \leq k$.
7) Consider the set of $R_1$ and the branch point of this set. Include the end-nodes that are omitted from this set at step 4. We initiate a branch trail originating form the branch point of this set. The link that connects the end-nodes in $R_1$ is the first link of this branch trail. Extend this branch trail to the opposite direction of the branch point as long as trail reaches to the edge of the branch. When there are multiple options, randomly pick one of the links to extend the branch trail.
8) Using *Lemma 5*, we can define the branch point as the starting point of this trail. This point behaves like the complementary of the end-nodes combined at this branch point. For example, if the combined end-nodes are $S_i \oplus T_j$, then the branch point would be seen as $T_i \oplus S_j$ over the branch trail.
9) Place the end-nodes over this trail using *Lemmas 1 to 4* as in step 3.
10) Repeat step 4 and 5. $R_1 = \{R_{1,1}, R_{1,2}, ..., R_{1,k_1}\}$, where $k_1$ is the number branch points over the first branch trail.
11) Return the step 6 iteratively as long as the all of the subbranches of the first branch is explored and each end-node is placed over a branch trail as a single entity. That will make sure that every end-node spanned by the first point is able to receive their parity data form the tree.
12) Pass to the next branch over the truck trail and return to step 7.

At the end, all of the end-nodes in CPP tree topology will be shown as a single entity in one of the linear 1+N coding trails, which makes them protected against single link failures. To clarify the steps shown above, an example is provided below.

*1) Example 2:* In the Example 1, a tree structure is partially converted to a linear 1+N coding trail. We proceed from the Example 1. At Fig. 12(b), it is shown that all of the end-nodes except $S'_3, T'_6, T'_5$ are placed over the linear 1+N coding trail, which enables them to code and encode their parity data over this trail. However, in order to protect the rest of the end-nodes, we need a branch trail originating from the branch point $D$. As shown in Fig. 12(d), the branch point is considered as $T'_3 \oplus S'_6 \oplus S'_5$ replacing the rest of the trail. According to *Lemma 5*, these three end-nodes can be shown as separate end-nodes as $T'_3, S'_6, S'_5$ over the branch trail. In addition, we include a new connection between $S7 - T7$ that is bounded inside this branch trail. The branch trail is extended as defined in step 2 in the previous section. The end-nodes that are spanned by this branch are placed over a new linear coding trail. This trail is depicted in Fig. 12(e). The introduction of a new connection does not affect the coding operations in the rest of the networks since the input and output signals at the branch point are the same. The end-nodes that are shown as single entities are protected by this coding trail. As in the truck trail, there is a branch point $E'$ that combines multiple end-nodes over the branch trail. It is required to go one more level down and generate a sub-branch trail to cover these end-nodes as well. This sub-branch trail is shown in Fig. 12(f). The operation is stopped when all of the end-nodes are place over a linear 1+N coding trail.

### D. Extensions on the Coding Group Selections

The link-disjointness rules in forming coding groups given in the previous section are sufficient to satisfy the encoding and decoding inside the network. There is still room to improve in terms of capacity efficiency without impairing the decodability of the coding structure. The first rule of link-disjointness is a necessary condition for decodability. However, the second rule can be modified to allow sharing of a common link by the primary path of a connection and the protection paths of other connections in the same coding group. The second rule will be

- Their primary paths are also link-disjoint with the primary paths of the connections in the same coding group.

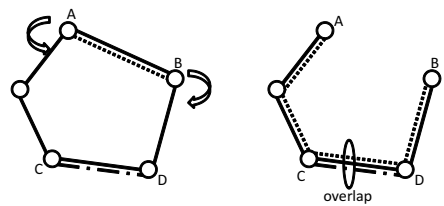

Fig. 13. Overlap of the primary and protection paths in cycle removal procedure

In this mode of operation, if the common link shared by one primary path and one or more protection paths fails, then the end-nodes of the failed protection paths can derive the failure over their protection paths by comparing the data from the primary paths with the data from the protection paths. After that the transmission over the failed protection paths will be temporarily terminated via the end-nodes of these protection paths. Otherwise, symmetricity is broken for more than one connection on the protection topology and decoding structure breaks down. In other words, the failed protection paths need to stop poisoning other protection paths because there are no antidotes.

Cycle elimination procedure for this mode of operation is not straightforward as it is in the previous case. Previously, the data over the longest link of a cycle is coded with the data over the rest of the cycle and that link is released from the coding topology. However, that may not be possible when the primary paths and protection paths of different connections share a common link. In that case, the protection path of a connection can overlap with its own primary path if it is rerouted over the rest of the cycle. It is depicted in Fig. 13. The straight lines are a portion of the protection topology. The direct dashed link between $A - B$ is a portion of the protection path of an arbitrary connection. The direct dashed link between $C - D$ is a portion of the primary path of the same arbitrary connection. Previously, the primary paths and the protection topology were link-disjoint. However, in this mode they can share some links. If the protection data over link $A - B$ is rerouted and coded with the data over the rest of the cycle then the primary and protection paths share a common link, which makes recovery impossible against that link failure.

To preserve the link-disjointness criterion between the primary and protection paths of the same connection, a new cycle elimination procedure is proposed.

1) Select the longest link on the cycle. Remove this link and code the data on it with the data over the rest of the cycle. Check if breaks down the link-disjointness between the primary and protection paths of each connection.
2) If so, select the next longest link until you find a link whose removal does not affect the link-disjointness criterion
3) If there is no such a link, look for a separation point on the cycle. A separation point on the cycle is a node whose incident (on-cycle) links carry no mutual data. In other words, this node is the end-node of the data on both of its incident links. If there is such a separation point, this cyclic structure can be considered as a tree structure and preserves the coding structure.
4) If there is no separation point on the cycle, then recalculate the route the portions of the protection paths that causes the conflict between link-disjointness and cyclic property.
5) If no solution is found then remove the connections which cause the conflict from the coding group. Protect these connections by 1+1 APS.

## IV. THE ILP FORMULATION

We developed an ILP formulation to find the optimal SPP solution for a given set of traffic scenarios and networks. We also developed an ILP formulation to convert this SPP solution into a viable CPP solution. The ILP formulation is varied to include both the wavelength continuity constraint and its absence in order to cover different types of optical networks.

As stated in [16], the problem of joint path routing and wavelength assignment is a very complex problem. Therefore we developed a suboptimal ILP formulation for the SPP algorithm. However, the ILP formulation of conversion from the SPP to the CPP algorithm is optimal. We input a set of possible paths for the connections and run the optimal wavelength assignment and sharing algorithm. Due to space limitations, the ILP formulation of SPP is not discussed here.

The ILP formulation of CPP has the following set of input parameters (When it is stated "equals 1 if A is true," it simultaneously means "equals 0 if A is not true")

- $G(V, E)$: The network graph
- $N$: Enumerated list of bidirectional connections
- $c_e$: Cost of each link
- $d_e(i)$: Equals 1 if the protection path of connection $i$ traverses over link $e$, is acquired from the solution of SPP
- $m(i, j)$: Equals 1 if the primary paths of connection $i$ and connection $j$ are link-disjoint, is acquired from the solution of SPP.

In addition to the input parameters, there are a number of binary variables

- $n(i, j)$: Equals 1 if connection $i$ and connection $j$ are in the same coding group
- $r_e(i, j)$: Equals 1 if connection $i$ and connection $j$ are coded together over link $e$
- $s_e(i)$: Equals 1 if protection capacity of connection $i$ over link $e$ can be saved with coding

Both $n(i, j)$ and $m(i, j)$ are defined in a way that $i \leq j$. The objective function is

$$\min \sum_{e \in E} \sum_{i \in N} c_e \times (d_e(i) - s_e(i)) \qquad (1)$$

subject to the following constraints

$$\begin{aligned} r_e(i,j) &\leq n(i,j), &\forall i,j \in N, i<j, \forall e \in E, &\quad (2)\\ r_e(i,j) &\leq d_e(i), &\forall i,j \in N, i<j, \forall e \in E, &\quad (3)\\ r_e(i,j) &\leq d_e(j), &\forall i,j \in N, i<j, \forall e \in E, &\quad (4)\\ n(i,j) &\leq m(i,j), &\forall i,j \in N, i<j. &\quad (5) \end{aligned}$$

Inequality (2) ensures that if two connections are coded over a link, then they must be in the same coding group. Inequalities (3) and (4) ensure that if two connections are coded over a link their protection paths must traverse over that link. Inequality (5) makes sure only link-disjoint connections can be in the same coding group. In addition,

$$n(i,j) \geq n(i,k) + n(k,i) + n(j,k) + n(k,j) - 1 \qquad (6)$$

$\forall i, j, k \in N, i < j$. Inequality (6) ensures that if connection $k$ is in the same coding group with connection $i$ and $j$, then connection $i$ and $j$ are also in the same coding group. The inequality

$$s_e(i) \leq \sum_{1 \leq k < i} r_e(k,i), \qquad \forall i,k \in N, k<i, \forall e \in E \quad (7)$$

calculates the savings due to the coding operation. When multiple protection paths are coded over the same link, only the one with the smallest index is accounted for the used capacity. Others save capacity by coding over the smallest indexed path.

ILP formulation for the *p*-cycle approach is adapted from [17]. Simulations are based on cycle exclusion-based ILP for spare capacity placement [17].

## V. RESTORATION TIME AND SIGNALING

In this section, we conduct a qualitative and a quantitative analysis in terms of restoration time of the SPP, CPP, and the *p*-cycle approaches. The analysis is extended to cover both opaque and transparent optical networks [18]. The developments in the optical XOR operations [19] allow coded shared protection to be applicable in all-optical networks [18]. Wavelength assignment of CPP is trivial after converting the wavelength assignment solution of SPP because CPP is inherently suited to the wavelength continuity constraint. With this constraint, protection paths in the same coding group make use of the same wavelength throughout the network. SPP [4] is proposed for all-optical networks but some shared path protection techniques, such as [7], and the *p*-cycle techniques make use of "optical-electrical-optical"(o-e-o) conversion at intermediate nodes.

Protection in CPP is a proactive mechanism because the second (protection) copy of any data is generated and transmitted by the source node to the destination node after a fixed time delay. An advantage of proactive protection mechanism is the continuous operation over protection paths which means there is no need to configure and test an OXC after any failure. OXC configuration and testing is the main source of delay in routing based protection mechanisms [20]. In addition, this proactive mechanism eliminates the need of complex signaling and assures transmission integrity because the operations are all automatic. As stated in [20], transmission integrity can be the main problem in configuring protection paths and routing data over these paths in optical networks. This claim is supported by the stability concerns cited in [6].

Despite the fact that CPP is a proactive mechanism, it can utilize the signaling capabilities of opaque optical networks to fasten the recovery process. In some cases, a synchronization mechanism with different data streams can neutralize the time savings of CPP. For that purpose, we propose a two-tier protection mechanism available for opaque networks. Transparent networks need to stick with the proactive mechanism due to the weak signaling capability of all-optical networks. In the first step, protection prompts as it is synchronized. As a second step, when the end-nodes receive the error signals, they add one bit control message to the data signals, which transform them into "ambulance" signals. The "ambulance" signals skip the buffers and are coded and encoded with data streams consisting of all zeros. The second step is similar to SPP but there is no need to configure the cross-connects. As a result, CPP is faster and more stable even in the worst case. The restoration time of the first part of the operation is

$$RT_{\text{CPP1}} = d_{sd} + h_b \times M + S,$$

where $d_{sd}$ is the propagation delay from node $s$ to node $d$ and $h_b$ is the number of hops in the protection path between $d$ and $s$. The symbol $M$ denotes the node processing time and varies on the type of optical network. It is taken 0.3 ms in [12] and 10 $\mu$s in [2]. The symbol $S$ represents the delay due to synchronization. The restoration time formulation of the second step is

$$RT_{\text{CPP2}} = F + 2 \times d_{sd} + (h_{is} + 1) \times M + (h_b + 1) \times M,$$

where $F$ is the failure detection time and $h_{is}$ is number of nodes between node $i$ and node $s$. The exact formulation of CPP for opaque optical networks is

$$RT_{\text{CPP}} = \min(RT_{\text{CPP1}}, RT_{\text{CPP2}}).$$

The recovery process in SPP starts with failure detection. Failure notification is required before end-nodes switch the traffic from primary to protection paths. Intermediate nodes configure themselves after they receive error state signals. In the protection switching step, some researchers claim that nodes in the protection path configure the OXCs simultaneously which leads to significant restoration time savings [16]. Error state signals should be transmitted over a specialized control plane to notify every node to enable simultaneous configuration of cross-connects over the protection path. As a tradeoff, this incurs high signaling complexity throughout the network. The restoration time formulation of SPP is under discussion, e.g., the results of some of the formulations [4], [16] do not match the numerical results in [2]. The OXC configuration time is stated to possibly be about 10 ms, but it is also reported to be as much as one second [6]. In addition, in [21] it is pointed out that an extra 40-80 ms is required only for signaling and reconfiguration, such as uploading maps. This means the OXC configuration time is not the only source of delay in SPP. Keeping the ambiguity in mind, we adopt the formula in [16] for the quantitative analysis of restoration time in SPP, assuming a separate packet-based control plane exists and has the same topology with the network of interest. The symbol $X$ represents the OXC configuration and test time, so that

$$RT_{\text{SPP1}} = F + 2 \times d_{sd} + (h_{si} + 1) \times M + X + (h_b + 1) \times M.$$

If a specialized control plane does not exist, in other words if in-band signaling is employed, then the OXCs cannot transmit the control message before they reconfigures themselves. This leads to higher restoration time due to the reconfiguration of OXCs in series. The formula for this case is adopted from [4]

$$\begin{aligned} RT_{\text{SPP2}} =& F + d_{sd} + (h_{is} + 1) \times M + (h_b + 1) \times X \\ &+ 2 \times d_{sd} + 2 \times (h_b + 1) \times M. \end{aligned}$$

TABLE I
SIMULATION RESULTS OF COST 239 NETWORK

| COST 239 Network, 11 nodes, 26 spans | | | | | | |
|---|---|---|---|---|---|---|
| Scheme | SCP | ESCP | RT for different X values (ms) | | | |
| | | | 0.5ms | 1ms | 5ms | 10ms |
| CPP | 72.71% | 0% | 11.57 | 11.57 | 11.57 | 11.57 |
| SPP1 | 64.67% | 0% | 17.86 | 18.36 | 22.36 | 27.36 |
| SPP2 | 64.67% | 0% | 28.5 | 31.1 | 51.1 | 76.1 |
| $p$-cycle | 44.82% | 40-60% | 25.31 | 25.81 | 29.81 | 34.81 |

The $p$-cycle approach is fundamentally a mixture of link protection technique and ring-type protection technique. This approach generally results in lower restoration time than SPP since the operations are local. *P*-cycle offers pre-connected OXCs around the cycle, so it minimizes the number of reconfigurations of OXCs after a failure. Only the end-nodes of the failure need to switch the traffic. However, an efficient *p*-cycle consists of many nodes, and traverses a long distance. This can add significant propagation and node processing delays in relatively large networks, such as the U.S. long-haul network. We employ the formula in [22] to calculate the restoration time of the *p*-cycle technique. This formula is modified to include the propagation delay after the source-end node of the failed link switches to the *p*-cycle until destination-end node of the failed link receives new data packets. In this paper, $M$ is taken as 0.3 ms [23], because the *p*-cycle uses o-e-o conversions. The parameter $d$ is the longest propagation delay between any two nodes in a *p*-cycle and $h$ is the number of nodes in a cycle. Then,

$$RT_{p-\text{cycle}} = F + X + h \times M + d.$$

Numerical results of worst case restoration time for the three techniques and a quantitative analysis are provided in the next section.

## VI. SIMULATION RESULTS

In this section, we will present simulation results for link failure recovery techniques previously discussed, in terms of their spare capacity requirements and worst case restoration time. In order to conduct a fair comparison between protection schemes, we input the same set of possible routing scenarios to ILP formulations of the SPP and the *p*-cycle approach. Since the CPP solutions are based on the SPP solutions, they also utilize the same set of routing scenarios.

The first network studied is the European COST 239 [24] network whose topology is given in Figure 14. In Fig. 14 and Fig. 15, the numbers associated with the nodes represent a node index, while the numbers associated with the edges correspond to the distance (cost) of the edge. The distances are useful to calculate the propagation delays. The traffic demand is uniform. SCP represents spare capacity percentage and explained in [11]. We provide the SCP values without the wavelength continuity constraint and restoration time results for the three schemes in Table I. In the third column, ESCP means the extra spare capacity percentage required to satisfy the wavelength continuity constraint [25].

Second network studied is the NSFNET network [13], similar to the U.S long-haul network [11]. Again, the traffic scenario is uniform. We provide the SCP values and restoration time results for the three schemes in Table II.

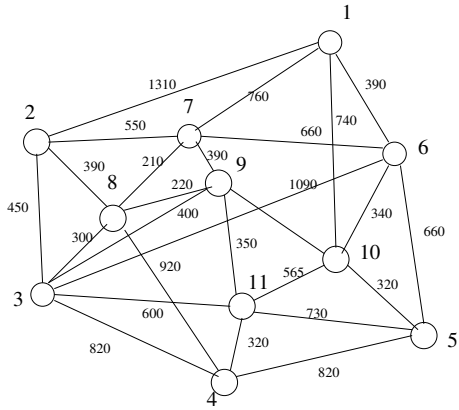

Fig. 14. European COST 239 network.

TABLE II
SIMULATION RESULTS OF NSFNET NETWORK

| NSFNET Network, 14 nodes, 21 spans | | | | | | |
|---|---|---|---|---|---|---|
| Scheme | SCP | ESCP | RT for different X values (ms) | | | |
| | | | 0.5ms | 1ms | 5ms | 10ms |
| CPP | 95.26% | 0% | 34.65 | 34.65 | 34.65 | 34.65 |
| SPP1 | 88.41% | 1.43% | 51.54 | 52.04 | 56.04 | 61.04 |
| SPP2 | 88.41% | 1.43% | 79.01 | 81.51 | 101.5 | 126.5 |
| $p$-cycle | 84.51% | 40-60% | 76.43 | 76.93 | 80.93 | 85.93 |

As seen from the results, converting the SPP solution to CPP results in approximately 6-7% extra spare capacity percentage. On the other hand, the restoration speed increases three times over SPP2 when in-band signaling is used and increases two times over SPP1 when there is a separate control plane for SPP scheme. The restoration time of SPP increases as the expected time of OXC configuration and test increases. Realistically, in some cases it may take seconds. The $p$-cycle technique results in lower SCP than CPP without the wavelength continuity constraint. Under the wavelength continuity constraint, CPP is as capacity efficient as the $p$-cycle technique for the COST 239 network and is more capacity efficient than $p$-cycle technique for the NSFNET network. It is observed that capacity efficiency of the $p$-cycle technique vanishes while going towards more sparse networks. The CPP is at least twice as fast as $p$-cycle technique.

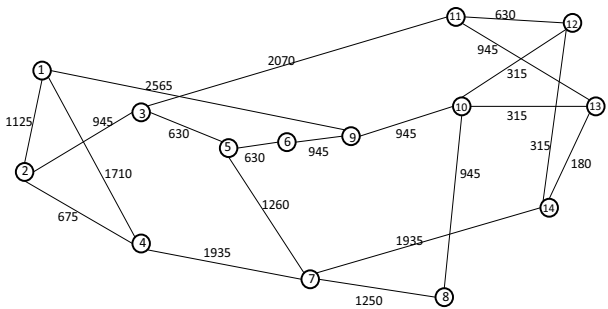

Fig. 15. NSFNET network.

## VII. CONCLUSION

In this paper, we introduced a proactive network restoration technique we call Coded Path Protection (CPP). The technique makes use of symmetric transmission over protection paths and link-disjointness among the connections in the same coding group. We modified the coding structure and leveraged its flexibility to convert sharing structure of a typical solution of SPP into a coding structure of CPP in a simple manner. With this approach, it is possible to quickly achieve close to optimal solutions. As a result of this operation, the CPP algorithm has the following properties.

- The restoration speed is 2 to 3 times faster than SPP and the $p$-cycle technique
- Full transmission integrity and stability
- Low signaling complexity
- Protection is independent of any single link failure scenario
- Simulation complexity significantly reduced over generalized 1+N coding
- Lower spare capacity than $p$-cycle under wavelength continuity constraint

with the tradeoff of

- At most 6-7 % extra spare capacity over SPP
- Lower capacity efficiency than $p$-cycle technique in dense networks
- Additional synchronization and buffering

Although the capacity placement algorithm for SPP employed in this paper is not optimal, this does not affect the optimality of the conversion algorithm.